\documentstyle[11pt,twoside]{article}
\begin{document}\hbadness=10000
\pagestyle{myheadings}\thispagestyle{empty}
\markboth{J. Rafelski and J. Letessier}
{Hadron Freeze-out and QGP  Hadronization}
\title{Hadron Freeze-out and  QGP Hadronization}
\author{Johann Rafelski and Jean Letessier\\ $ $\\
Department of Physics, University of Arizona, Tucson, AZ 85721\\
and\\
Laboratoire de Physique Th\'eorique et Hautes Energies\\
Universit\'e Paris 7, 2 place Jussieu, F--75251 Cedex 05.
}
\date{February 16, 1999}
\maketitle
\begin{abstract}
Abundances and $m_\bot$-spectra 
of strange and other hadronic particles 
emanating from central 158--200~A~GeV
reactions between nuclei are found to originate from a thermally 
equilibrated, deconfined source in  chemical non-equilibrium. 
Physical freeze-out parameters and physical conditions 
(pressure, specific energy, entropy,
and strangeness) are evaluated. Five properties
of the source we obtain are as expected 
for direct hadron emission (hadronization) 
from a deconfined quark-gluon plasma phase.
\end{abstract}
\begin{center} 
\end{center}

\section{Introduction}
Quark-gluon plasma (QGP) is, by the meaning  
of these words, a thermally equilibrated  
state  consisting of mobile, color charged 
quarks and gluons.  Thermal equilibrium 
is established faster than the particle abundance 
(chemical) equilibrium and thus in general
quark and gluon abundances in QGP can 
differ from their equilibrium Stefan-Boltzmann limit. 
This in turn impacts the hadronic particle production yields,
and as we shall see, the chemical non-equilibrium is a key 
ingredient in the successful data analysis of experiments 
performed at the  CERN-SPS in the past decade. We address here
results of 200~A~GeV Sulphur (S)  beam interactions
with laboratory stationary targets, such as
Gold (Au), Tungsten (W) or Lead (Pb) 
nuclei and, Pb--Pb collisions at 158~A~GeV. In these interactions 
the nominal center of momentum (CM)  available energy 
is  8.6--9.2 GeV per participating nucleon.

Considerable refinement of data analysis  has occurred since
last comprehensive review of the field \cite{Sol97}. Our present 
work includes in particular the following:
\begin{enumerate}
\item 
We considered aside of strange also 
light quark ($q=u,d$) chemical non-equilibrium
abundance \cite{LRa99} and introduce along 
with the  statistical strangeness 
non-equilibrium parameter $\gamma_s$, its light quark
analogue $\gamma_{q}$.
\item 
Coulomb distortion of the strange quark
phase space has been understood \cite{LRPb98}.
\end{enumerate}
Incorporating these developments, 
we accurately describe abundances of strange as well as non-strange
hadrons, both at central rapidity and in $4\pi$-acceptance. We
are thus combining  in the present 
analysis the strangeness diagnostic tools
of dense matter with the notion of entropy 
enhancement in deconfined phase~\cite{Let93}.

As particle emerge from the QGP hadronization, not only their abundance
but also their spectral shape is of interest. Our analysis  
considers the impact of explosive radial flow on
the spectra of particles at high $m_\bot$. This 
contributes significant information about 
the fireball dynamics and the possibly deconfined 
nature of the hadron source.

\section{The Coulomb effect in the quark fireball}\label{Coulombsec}
The diverse statistical chemical parameters that we need to consider will 
in a self-explanatory way be now introduced, considering the concept of 
strangeness balance: since strangeness can only be made and destroyed by 
hadronic interactions in pairs, the net abundance in the hadronic fireball
must vanish. We consider a hot gas of  free quarks, and 
evaluate the difference between strange and anti-strange quark numbers
(net strangeness). The  Coulomb potential  originating in the  initial 
proton abundance distorts slightly the Fermi particle distributions:
 strange quarks (negative charge) are attracted, while 
anti-strange quarks repelled. Allowing for
this slowly changing potential akin to the relativistic Thomas-Fermi phase space
occupancy model at finite temperature, the energy of a quark depends on both 
the momentum and the Coulomb potential $V$:
\begin{equation}\label{EpV}
E_p=\sqrt{m^2+p^2}-\frac 13 V\,.
\end{equation}
It is helpful to see here in first instance the potential $V$ 
as a square well, Eq.\,(\ref{EpV})  within the volume of
interest. Counting the states in the fireball we obtain:
\begin{eqnarray}
\langle N_s-N_{\bar s}\rangle =\int_{R_{\rm f}} 
        g_s\frac{d^3rd^3p}{(2\pi)^3}
\left[
 \frac1{1+\gamma_s^{-1}\lambda_s^{\,-1}e^{(\sqrt{m^2+p^2}-\frac13 V)/T}}
-\frac1{1+\gamma_s^{-1}\lambda_se^{(\sqrt{m^2+p^2}+\frac13 V)/T}}\right]
  \,.\label{Nsls}
\end{eqnarray}
In Eq.\,(\ref{Nsls}) the subscript ${R_{\rm f}}$ on the spatial integral 
reminds us that only the classically
allowed region within the fireball is covered in the integration over the 
level density, $g_s(=6)$ is the quantum degeneracy of
strange quarks. The magnitude of the charge of strange quarks ($Q_s/|e|=1/3)$
is shown explicitly,  the Coulomb potential refers to a
negative integer probe charge. 

The fugacity $\lambda_s$ of strange quarks enters particles
and antiparticles with opposite power, while the occupancy parameter $\gamma_s$ 
enters both term with same power. For $\gamma_s<1$ 
one obtains a rather precise result 
for the range  of parameters of interest
to us (see below) considering the Boltzmann approximation:
\begin{equation}\label{NslsB}
\langle N_s-N_{\bar s}\rangle \simeq
\frac{\int_{R_{\rm f}} d^3r\left[\lambda_s e^{\frac V{3T}}
 -\lambda_s^{-1} e^{-\frac V{3T}}\right]}{\int_{R_{\rm f}} d^3r}
\, g_s\int\frac{d^3pd^3x}{(2\pi)^3}\gamma_s e^{-\sqrt{p^2+m^2}/T}\,.
\end{equation}
The meaning of the different factors is now 
evident. $\gamma_s$ controls overall 
abundance of strange quark pairs,  multiplying 
the usual Boltzmann thermal factor
while $\lambda_s$ controls the difference
between the number of strange and non-strange 
quarks. Since strangeness is produced 
as $s,\bar s$-pair, the value of $ \lambda_s$ 
fulfills the constraint
\begin{equation}\label{sbalance}
\int_{R_{\rm f}} d^3r\left[ \lambda_s e^{\frac V{3T}}
 -\lambda_s^{-1} e^{-\frac V{3T}}\right]=0\,,
\end{equation}
which is satisfied exactly both in the Boltzmann limit 
Eq.\,(\ref{NslsB}), and for the exact quantum
distribution Eq.\,(\ref{Nsls}), when:
\begin{equation}\label{tilams}\label{lamQ}
\tilde\lambda_s\equiv \lambda_s \lambda_{\rm Q}^{1/3}=1\,,\qquad
\lambda_{\rm Q}\equiv 
\frac{\int_{R_{\rm f}} d^3r e^{\frac V{T}} } {\int_{R_{\rm f}} d^3r}\,.
\end{equation}
$\lambda_{\rm Q}$ is not a fugacity that can 
be adjusted to satisfy a chemical condition,
its value is defined by the applicable Coulomb potential $V$. 
More generally, in order to 
account for the Coulomb effect, the quark fugacities 
within a deconfined region should be renamed 
as follows in order to absorb the 
Coulomb potential effect: 
\begin{eqnarray}\label{lamtil}\nonumber
&\lambda_s\to \tilde\lambda_s\equiv\lambda_s\lambda_Q^{1/3} \,, \,\ \quad& 
\lambda_d\to \tilde\lambda_d \equiv\lambda_d\lambda_Q^{1/3}\,, 
\quad 
\label{tildl}
\ \nonumber\\
&\lambda_u\to \tilde\lambda_u \equiv\lambda_u\lambda_Q^{-2/3}\,, \quad&
\lambda_q\to \tilde\lambda_q\equiv\sqrt{\tilde\lambda_u\tilde\lambda_d}
=\lambda_q\lambda_Q^{-1/6}\,.
\end{eqnarray}
Since $Q_d=Q_s=-1/3$, the first line is quite evident 
after the above strangeness  discussion, the
second follows with $Q_u=+2/3$. Note that 
for a negatively charged strange quark the 
tilded fugacity Eq.\,(\ref{tilams}) contains 
a factor with positive power $1/3$
but the  potential that enters the 
quantity $\lambda_{\rm Q}$ is negative, and thus 
$\tilde\lambda_s<\lambda_s$. Because Coulomb-effect acts in
opposite way on $u$ and $d$ quarks, its net impact on $\lambda_q$ is
smaller than on $\lambda_s$, and it also acts in the opposite way with
$ \tilde\lambda_q>\lambda_q$. To see the relevance 
of the tilde fugacities for light quarks, note that in order to obtain 
baryon density  in QGP one needs to use 
the tilde-quark fugacity to account for the Coulomb potential 
influence on the phase space.

It is somewhat unexpected 
that for the Pb--Pb fireball the Coulomb effect is at all 
relevant. Recall that for a uniform charge 
distribution within a radius $R_{\rm f}$ of charge $Z_{\rm f}$:
\begin{equation}
V=\left\{
\begin{array}{ll}\displaystyle
-\frac32 \frac{Z_{\rm f}e^2}{R_{\rm f}}
      \left[1-\frac13\left(\frac r{R_{\rm f}}\right)^2\right]\,,
          & \mbox{for}\quad r<R_{\rm f}\,;\\
-\displaystyle\frac{Z_{\rm f}e^2}{r}\,,& \mbox{for} \quad r>R_{\rm f}\,.
\end{array}
\right.
\end{equation}
Choosing $R_{\rm f}=8$\,fm, $T=140$\,MeV,
$m_s=200$\,MeV (value of $\gamma_s$ is practically irrelevant) we find as 
solution of Eq.\,(\ref{Nsls})  for $\langle N_s-N_{\bar s}\rangle=0$
for $Z_{\rm f}=150$ the value $\lambda_s=1.10$
(precisely: 1.0983; $\lambda_s=1.10$ corresponds to $R_{\rm f}=7.87$\,fm). 
We will see that both this values within the experimental precision arise
from study of particle abundances.  
For the S--W/Pb reactions this Coulomb effect is
practically negligible.

In the past we (and others) have disregarded 
in the description of the hadronic final state abundances
the electrical charges and interactions of the produced hadrons.
This is a correct way to analyze the chemical properties since,
as already mentioned, the quantity 
$\lambda_{\rm Q}$ is not a new fugacity: conservation
of flavor already assures charge conservation 
in chemical hadronic reactions, and use 
of $\lambda_i,\ i=u,d,s$ exhausts all available 
chemical balance conditions for the 
abundances of hadronic particles. As shown here, the
Coulomb deformation of the phase space in the QGP 
fireball makes it necessary to 
rethink the implications that the final state particle measured 
fugacities have on the fireball properties.

\section{Freeze-out of hadrons}
The production of hadrons from a QGP fireball  occurs mainly by way of quark 
coalescence  and gluon fragmentation, and there can 
be some quark fragmentation as well.
We will explicitly consider the recombination of quarks, but implicitly the 
gluon fragmentation is accounted for by our allowance for chemical 
nonequilibrium. The relative number of primary 
particles freezing out from a source is obtained 
noting that the  fugacity and phase space occupancy 
 of a composite hadronic  particle can be   expressed  
by its constituents and that the probability to find all 
$j$-components contained within  the $i$-th  emitted particle is:
\begin{equation}\label{abund}
N_i\propto \prod_{j\in i}\gamma_j\lambda_je^{-E_j/T}\,,
\qquad\lambda_i=\prod_{j\in i}\lambda_j\,,
\qquad \gamma_i=\prod_{j\in i}\gamma_j\,.
\end{equation}
Experimental data with
full phase  space coverage, or central rapidity region 
$|y-y_{\rm CM}|<0.5$, for $m_\bot>1.5$~GeV  
are considered; recall that  the energy of a hadron `i' is expressed by
the spectral parameters $m_\bot$ and $y$ as follows,
 \[
E_i=\sum E_j\,,\qquad
E_i=\sqrt{m_i^2+p^2}=\sqrt{m_i^2+p_\bot^2}\cosh (y-y_{\rm CM})\,, \] 
where $y_{\rm \small CM}$ is the center of momentum rapidity of the fireball formed 
by the colliding nuclei.

The yield of particles is  controlled by the
freeze-out temperature $T_{\rm f}$. This freeze-out temperature
is different from the $m_\bot$-spectral temperature $T_\bot$, which 
also comprises the effect of  collective matter flows originating in 
the explosive disintegration driven by the 
internal pressure of compressed hadronic matter. 
In order to model the flow and freeze-out 
of the fireball surface, one in general 
needs several new implicit and/or explicit parameters.  We therefore 
will make an effort to choose experimental variables (compatible particle
ratios) which are largely flow independent. This approach also 
diminishes the influence of 
heavy resonance population --- we include in Eq.(\ref{abund})
hadronic states up to $M=2$ GeV, and also include quantum statistical 
corrections, allowing for first Bose and Fermi distribution corrections in
the phase space content. It is hard to check if indeed we succeeded in 
eliminating the uncertainty about high mass hadron populations. 
As we shall see comparing descriptions which exclude flow with those
that include it, our approach is indeed 
largely flow-insensitive.

We consider  here a simple radial flow model, 
with freeze-out in CM frame at constant laboratory time, implying that
causally disconnected domains  of the dense matter fireball are
 synchronized at the  instant of the collision.
Within this approach \cite{Hei92}, the spectra and thus also multiplicities 
of particles emitted are obtained replacing
the Boltzmann exponential factor in Eq.(\ref{abund}), 
\begin{equation}\label{abundflow}
e^{-E_j/T}\to \frac1{2\pi}\int d\Omega_v
  \gamma_v(1+\vec v_{\rm c}\cdot \vec p_j/E_j)
  e^{-{{\gamma_vE_j}\over T}
    \left(1+\vec v_{\rm c}\cdot \vec p_j/E_j\right)} \,,
\end{equation}
where as usual $\gamma_v=1/{\sqrt{1-\vec v_{\rm c}^{\,2}}}$\,. 
Eq.\,(\ref{abundflow}) can  be intuitively obtained by a Lorentz
transformation between an observer on the surface of
the fireball, and one at rest in the general CM (laboratory) frame.
One common feature of all flow
scenarios is that, at sufficiently high $m_\bot$, the spectral temperature 
(inverse slope) $T_\bot$ can be derived from the freeze-out temperature 
$T_{\rm f}$ with the help of the Doppler formula:
\begin{equation}\label{doppler}
T_\bot=T_{\rm f}\gamma_v(1+ v_c)\,.
\end{equation}
In actual 
numerical work, we proceed as follows to account for the Doppler effect:
for a given pair of values $T_{\rm f}$ and $v_{\rm c}$, the resulting  
$m_\bot$ particle spectrum is obtained and analyzed using the spectral shape 
and procedure employed for the particular collision system 
by the experimental groups, yielding
the theoretical inverse slope `temperature' $T_{\bot}^j$\,.

Once the parameters  $T_{\rm f},\,\lambda_{q},\,\lambda_{s},\,
\gamma_{q},\,\gamma_{s}$ and $v_c$ have been determined studying   
available particle yields, and $m_\bot$ slopes, the entire phase space 
of particles produced is fully characterized within our elaborate statistical model. 
Our model is in fact just an elaboration of the original Fermi model \cite{Fer50}, 
in fact all we do is to allow hadronic particles to be produced in the manner
dictated by the phase space size of valance quarks. With the full understanding of 
the phase space of all hadrons, we can evaluate the 
physical properties of the system at freeze-out, such as, {\it e.g.}, 
energy and entropy per baryon, strangeness content. 

\section{Results of data analysis}
As noted our analysis requires that we form  particle abundance ratios between 
what we call compatible hadrons. 
We  considered for S--Au/W/Pb reactions 
18 data points listed in table~\ref{resultsw} 
(of which three comprise  results with $\Omega$'s). For Pb--Pb we address 
here 15 presently available particle yield ratios 
listed in table~\ref{resultpb}
(of which four  comprise  the $\Omega$'s). 
We believe to have included in our discussion most if not all 
particle multiplicity results available presently.
\begin{table}[tb]
\caption{\protect\label{resultsw}
Particle ratios studied in our analysis for  S--W/Pb/Au reactions: 
experimental results with  references and kinematic cuts are given, 
followed by columns showing results for the different strategies  of analysis B--F.
Asterisk~$^*$ means a predicted result (corresponding data is not 
fitted). Subscript $s$ implies forced strangeness conservation, subscript $v$ 
implies inclusion of collective flow. The experimental results considered are 
from:}
\vspace{0.1cm}
$^1$ {S.\,Abatzis {\it et al.}, WA85 Collaboration, {\it Heavy Ion Physics} {\bf 4}, 79 (1996).} \\
$^2$ {S.\,Abatzis {\it et al.}, WA85 Collaboration, {\it Phys.\,Lett.}\,B {\bf 347}, 158 (1995).} \\ 
$^3$ {S.\,Abatzis {\it et al.}, WA85 Collaboration, {\it Phys.\,Lett.}\,B {\bf 376}, 251 (1996).} \\ 
$^4$ {I.G.\,Bearden {\it et al.}, NA44 Collaboration, {\it Phys.\,Rev.}\,C {\bf 57}, 837 (1998).}  \\
$^5$ {D.\,R\"ohrich for the NA35 Collaboration, {\it Heavy Ion Physics} {\bf 4}, 71 (1996).} \\ 
$^6$ S--Ag value adopted here: {T.\,Alber {\it et al.}, NA35 Collaboration, 
{\it Eur.\,Phys.\,J.} C {\bf 2}, 643 (1998).}\\
$^7$ A. Iyono {\it et al.}, EMU05 Collaboration, {\it Nucl.\,Phys.\,}A {\bf 544},
455c (1992) and  Y. Takahashi {\it et al.}, EMU05 \hspace*{0.3cm}Collaboration, 
private communication.
\small\begin{center}
\begin{tabular}{|lclc|lllll|ll|}
\hline\hline    
 Ratios & $\!\!\!\!$Ref. & Cuts [GeV] & Exp.Data      & B        &C         &D         & D$_s$    & F    &D$_v$     & F$_v$ \\
\hline
${\Xi}/{\Lambda}$                           & 1 &   
$1.2<p_{\bot}<3$   &0.097 $\pm$ 0.006                & 0.16     & 0.11     & 0.099    &0.11      & 0.10 &0.11      &0.11   \\
${\overline{\Xi}}/{\bar\Lambda}$            & 1 &  
$1.2<p_{\bot}<3$ &0.23 $\pm$ 0.02                    & 0.38     & 0.23     & 0.22     &0.18      & 0.22 &0.23      &0.22  \\
${\bar\Lambda}/{\Lambda}$                   & 1 &
$1.2<p_{\bot}<3$ &0.196 $\pm$ 0.011                  & 0.20     & 0.20     & 0.203    &0.20      & 0.20 &0.20      &0.20  \\
${\overline{\Xi}}/{\Xi}$                    & 1 &    
$1.2<p_{\bot}<3$    &0.47 $\pm$ 0.06                 & 0.48     & 0.44     & 0.45     &0.33      & 0.44 &0.44      &0.43  \\
${\overline{\Omega}}/{\Omega}$              & 2 &
$p_{\bot}>1.6$&0.57 $\pm$ 0.41                       &1.18$^{*}$&0.96$^{*}$&1.01$^{*}$&0.55$^{*}$& 0.98 &1.09$^{*}$&1.05$^{*}$\\
$\Omega+\overline{\Omega}\over\Xi+\bar{\Xi}$& 2 &
{$p_{\bot}>1.6$}&0.80 $\pm$ 0.40                     &0.27$^{*}$&0.17$^{*}$&0.16$^{*}$&0.16$^{*}$& 0.16 &0.13$^{*}$&0.13$^{*}$  \\
${K^+}/{K^-}$                               & 1 & 
{$p_{\bot}>0.9$}         &1.67  $\pm$ 0.15           & 2.06     & 1.78     &  1.82    &1.43      & 1.80 &1.77      &1.75  \\
${K^0_{\rm s}}/\Lambda$                     & 3 & 
{$p_{\bot}>1$}      &1.43  $\pm$ 0.10                & 1.56     & 1.64     &  1.41    &1.25      & 1.41 &1.38      &1.39  \\
${K^0_{\rm s}}/\bar{\Lambda}$               & 3 &
{$p_{\bot}>1$}  &6.45  $\pm$ 0.61                    & 7.79     & 8.02     &  6.96    &6.18      & 6.96 &6.81      &6.86  \\
${K^0_{\rm s}}/\Lambda$                     & 1 & 
{$m_{\bot}>1.9$}    &0.22  $\pm$ 0.02                & 0.26     & 0.28     &  0.24    &0.24      & 0.24 &0.24      &0.24  \\
${K^0_{\rm s}}/\bar{\Lambda}$               & 1 &
{$m_{\bot}>1.9$}  &0.87  $\pm$ 0.09                  &1.30      & 1.38     &  1.15    &1.20      & 1.16 &1.18      &1.17  \\
${\Xi}/{\Lambda}$                           & 1 &
{$m_{\bot}>1.9$}       &0.17 $\pm$ 0.01              & 0.27     & 0.18     &  0.17    &0.18      & 0.17 &0.16      &0.17  \\
${\overline{\Xi}}/{\bar\Lambda}$            & 1 &
{$m_{\bot}>1.9$}&0.38 $\pm$ 0.04                     & 0.64     & 0.38     &  0.38    &0.30      & 0.37 &0.35      &0.35  \\
$\Omega+\overline{\Omega}\over\Xi+\bar{\Xi}$& 1 &
{$m_{\bot}>2.3$}&1.7 $\pm$ 0.9                       &0.98$^{*}$&0.59$^{*}$&0.58$^{*}$&0.52$^{*}$& 0.58 &0.72$^{*}$&0.75$^{*}$  \\
$p/{\bar p}$                                & 4 &
Mid-rapidity &11 $\pm$ 2\ \                          & 11.2     & 10.1     & 10.6     &7.96      & 10.5 &10.6      &10.4  \\
${\bar\Lambda}/{\bar p}$                    & 5 & 
4 $\pi$   &1.2 $\pm$ 0.3                             & 2.50     & 1.47     & 1.44     &1.15      & 1.43 &1.58      &1.66  \\
$h^-\over p-\bar p$                         & 6 &        
4 $\pi$   &4.3 $\pm$ 0.3                             & 4.4      & 4.2      & 4.1      & 3.6      & 4.1  &4.2       &4.2  \\
$h^+-h^-\over h^++h^-$                      & 7 &  
4 $\pi$   &0.124 $\pm$ 0.014                         & 0.11     & 0.10     & 0.103    &0.09      & 0.10 &0.12      &0.12  \\
\hline
 &                           & $\chi^2_{\rm T}$ &    & 264      & 30       & 6.5      & 38       & 12   &6.2       &11  \\
\hline\hline
\end{tabular}
\end{center}
 \end{table}
\begin{table}[tb]
\caption{\label{resultpb}
Particle ratios studied in our analysis for  Pb--Pb reactions: 
experimental results with  references and kinematic cuts are given, 
followed by columns showing results for the different strategies of analysis B--F.
Asterisk~$^*$ means a predicted result (corresponding data is not 
fitted or not available).  Subscript $s$ implies forced strangeness 
conservation, subscript $v$ 
implies inclusion of collective flow. The experimental results considered are 
from:}
\vspace*{0.1cm}
$^1$ {I.\,Kr\'alik, for the WA97 Collaboration, {\it Nucl. Phys.} A {\bf 638},115, (1998).}\\
$^2$ {G.J.\,Odyniec, for the NA49 Collaboration, {\it J. Phys.} G {\bf 23}, 1827 (1997).}\\
$^3$ {P.G.\,Jones, for the NA49 Collaboration, {\it Nucl. Phys.} A {\bf 610}, 188c (1996).}\\
$^4$ {F.\,P\"uhlhofer, for the NA49 Collaboration, 
{\it Nucl. Phys.} A {\bf 638}, 431,(1998).}\\
$^5$ {C.\,Bormann, for the NA49 Collaboration, {\it J. Phys.} G {\bf 23}, 1817 (1997).}\\
$^6$ {G.J.\,Odyniec,  {\it Nucl. Phys.} A  {\bf 638}, 135, (1998).}\\
$^7$ {D. R\"ohrig,  for the NA49 Collaboration, 
``Recent results from NA49 experiment on 
Pb--Pb collisions at \linebreak \hspace*{0.25cm}158 A GeV'',
see  Fig. 4, in proc. of EPS-HEP Conference, Jerusalem, Aug. 19-26, 1997.}\\
$^8$ {A.K.\,Holme, for the WA97 Collaboration, {\it J. Phys.} G {\bf 23}, 1851 (1997).}
\small\begin{center}
\begin{tabular}{|lclc|lllll|ll|}
\hline\hline
 Ratios & $\!\!\!\!$Ref. & Cuts [GeV] & Exp.Data                             & B       &C        &D        & D$_s$   & F       &D$_v$    & F$_v$ \\
\hline
${\Xi}/{\Lambda}$ &{1} &$p_{\bot}>0.7$&0.099 $\pm$ 0.008                    &0.138    &  0.093  &  0.095  &  0.098  &  0.107  & 0.102   & 0.110\\
${\overline{\Xi}}/{\bar\Lambda}$ &{1} &$p_{\bot}>0.7$&0.203 $\pm$ 0.024     &0.322    &  0.198  &  0.206  &  0.215  &  0.216  & 0.210   & 0.195\\
${\bar\Lambda}/{\Lambda}$  &{1} &$p_{\bot}>0.7$&0.124 $\pm$ 0.013           &0.100    &  0.121  &  0.120  &  0.119  &  0.121  & 0.123   & 0.128\\
${\overline{\Xi}}/{\Xi}$  &{1} &$p_{\bot}>0.7$&0.255 $\pm$ 0.025            &0.232    &  0.258  &  0.260  &  0.263  &  0.246  & 0.252   & 0.225\\
$(\Xi+\bar{\Xi})\over(\Lambda+\bar{\Lambda})$&{2} &$p_{\bot}>1.$ &0.13 $\pm$ 0.03 
                                                                            &0.169    &  0.114  &  0.118  &  0.122  &  0.120  & 0.123   & 0.121\\
${K^0_{\rm s}}/\phi$   &{3,4} &  &11.9 $\pm$ 1.5\ \                         &6.3      &  10.4   &  9.89   &  9.69   &  16.1   & 12.9    & 15.1\\
${K^+}/{K^-}$         &{5} &              &1.80  $\pm$ 0.10                 &1.96     &  1.75   &  1.76   &  1.73   &  1.62   & 1.87    & 1.56\\
$p/{\bar p}$     &{6} &            &18.1 $\pm$4.\ \ \ \                     &22.0     &  17.1   &  17.3   &  17.9   &  16.7   & 17.4    & 15.3\\
${\bar\Lambda}/{\bar p}$     &{7} &              &3. $\pm$ 1.               &3.02     &  2.91   &  2.68   &  3.45   &  0.65   & 2.02    & 1.29\\
${K^0_{\rm s}}$/B       &{3} &              &0.183 $\pm$ 0.027              &0.305    &  0.224  &  0.194  &  0.167  &  0.242  & 0.201   & 0.281\\
${h^-}$/B                 &{3} &              &1.83 $\pm $ 0.2\ \           &1.47     &  1.59   &  1.80   &  1.86   &  1.27   & 1.83    & 1.55\\
${\Omega}/{\Xi}$      &{1} &$p_{\bot}>0.7$&0.192 $\pm$ 0.024                &0.119$^*$&0.080$^*$&0.078$^*$&0.080$^*$&  0.192  &0.077$^*$& 0.190\\
${\overline{\Omega}}/{\overline{\Xi}}$  &{8} &$p_{\bot}>0.7$&0.27 $\pm$ 0.06&0.28$^*$ &0.17$^*$ &0.17$^*$ &0.18$^*$ &  0.40   &0.18$^*$ & 0.40\\
${\overline{\Omega}}/{\Omega}$  &{1} &$p_{\bot}>0.7$&0.38 $\pm$ 0.10        &0.55$^*$ &0.56$^*$ &0.57$^*$ &0.59$^*$ &  0.51   &0.60$^*$ & 0.47\\
$(\Omega+\overline{\Omega})\over(\Xi+\bar{\Xi})$&{8} &$p_{\bot}>0.7$&0.20 $\pm$ 0.03
                                                                            &0.15$^*$ &0.10$^*$ &0.10$^*$ &0.10$^*$ &  0.23   &0.10$^*$ & 0.23\\
\hline
& & $\chi^2_{\rm T}$ &                                                      & 88      & 24      & 1.6     & 2.7     & 19      & 1.5 & 18\\
\hline\hline
\end{tabular}
\end{center}
\end{table}

The theoretical particle yield results  
shown columns in tables  \ref{resultsw} and \ref{resultpb} 
are obtained looking for a set of physical parameters
 which will minimize the difference between theory and 
experiment.  The resulting total error of the ratios $R$ is
shown at the bottom of tables~\ref{resultsw}~and~\ref{resultpb}:
\begin{equation}
\chi^2_{\rm T}=\sum_j\left(
 \frac{R_{\rm th}^j-R_{\rm exp}^j}{{\Delta R _{\rm exp}^j}}
   \right)^2\,.
\end{equation}
It is a non-trivial matter  to determine the confidence level that
goes with the different data analysis approaches since some of the 
results considered are partially redundant, and a few
data points can be obtained from others by algebraic relations 
arising in terms of their 
theoretical definitions; there are two types of such relations:
\begin{equation}
\frac{\Omega+\overline{\Omega}}{\Xi+\bar{\Xi}}
=\frac{\overline\Omega}{\Xi}\cdot\,
    \frac{1+\overline\Omega/\Omega}{1+\overline\Xi/\Xi}\,,
\qquad 
\frac{\overline\Lambda}{\Lambda}=\frac{\overline\Lambda}{\overline\Xi}\cdot 
  \frac{\overline\Xi}{\Xi}\cdot \frac{\Xi}{\Lambda}\,.
\end{equation}
However, due to smallness of the total 
error found for some of the approaches it is clear without 
detailed analysis that only these are statistically significant.

In addition to the abundance data, we also 
explored the transverse mass $m_\bot$-spectra
when the collective flow velocity was allowed 
in the description, and the bottom line of 
tables~\ref{resultsw}~and~\ref{resultpb} 
in columns D$_v$, F$_v$ includes in these cases
the error found in the inverse slope parameter of the spectra.
The procedure we used is as follows: 
since within the error the high $m_\bot$ strange (anti)baryon
inverse slopes are within error, 
overlapping we decided to consider just one `mean' 
experimental  value $T_{\bot}=235\pm10$ for 
S--induced reactions and $T_{\bot}=265\pm15$
for Pb--induced reactions. Thus we add one 
experimental value and one parameter, without
changing the number of degrees of freedom. 
Once we find values of $T_{\rm f}$
and $v_{\rm c}$, we  evaluate the slopes of 
the theoretical spectra. The resulting theoretical
$T_{\rm th}^j$ values are  in remarkable 
agreement with experimental $T_{\bot}^j$, well beyond
what we expected, as is shown in 
table~\ref{Tetrange}. An exception is 
the fully strange $\Omega+\overline\Omega$ spectrum.  
We note in passing that when $v_c$ was introduced we found
little additional correlation between 
now 6 theoretical parameters. The collective 
flow velocity is a truly new degree of freedom  
and it helps to attain a  more consistent description of  
the experimental data available.
\begin{table}[ht]
\caption{\label{Tetrange}
Particle spectra inverse slopes: theoretical values $T_{\rm th}$ are obtained 
imitating the  experimental procedure from the $T_f,\,v_c$-parameters.
Top portion:  S--W  experimental $T_{\bot}$ from experiment WA85 for kaons, lambdas  
and cascades;
bottom portion: experimental Pb--Pb $T_{\bot}$ from experiment NA49 for 
kaons and from experiment WA97 for baryons.
The experimental results are  from:}
\hspace*{0.3cm}D. Evans, for the WA85-collaboration, APH N.S., Heavy Ion Physics {\bf 4},
79 (1996).\\
\hspace*{0.3cm}E. Andersen {\it et al.}, WA97-collaboration,  
{\it Phys. Lett.} B {\bf 433}, 209, (1998).\\
\hspace*{0.3cm}S. Margetis, for the NA49-collaboration, 
J. Physics G, Nucl. and Part. Phys. {\bf 25}, 189 (1999).
\begin{center}
\begin{tabular}{|l|cccccc|}
\hline\hline
&$T^{{\rm K}^0}$ &$T^\Lambda$& $T^{\overline\Lambda}$& $T^\Xi$
 &$T^{\overline\Xi}$& $T^{\Omega+\overline\Omega}$ \\
\hline
                 $T_{\bot}$ \ [MeV]
		     & 219 $\pm$  5
                 & 233 $\pm$  3
                 & 232 $\pm$  7 
                 & 244 $\pm$  12
                 & 238 $\pm$  16  
                 & ---         \\
                 $T_{\rm th}$ [MeV] 
		     &  215 
                 &  236
                 &  236 
                 &  246 
                 &  246    
                 &  260       \\
\hline
                 $T_{\bot}$ \ [MeV]
		     & 223 $\pm$  13
                 & 291 $\pm$  18
                 & 280 $\pm$  20 
                 & 289 $\pm$  12
                 & 269 $\pm$  22  
                 & 237 $\pm$  24         \\
                 $T_{\rm th}$ [MeV] 
		     &  241 
                 &  280
                 &  280 
                 &  298 
                 &  298    
                 &  335       \\
\hline\hline
\end{tabular}
\end{center}
\end{table}

Although it is clear that one should be using a full-fledged model such as $D_v$,
we address also cases B and C. The reason for this arises from  our desire to 
demonstrate the empirical need for chemical non-equilibrium: in the approach B, 
complete chemical equilibrium $\gamma_i=1$ is assumed. 
As we  see in tables~\ref{resultsw}~and~\ref{resultpb}
this approach has rather large error. Despite this the results 
in column B in  tables~\ref{resultsw}~and~\ref{resultpb}
are often compared favorably with experiment, indeed this result 
can be presented quite convincingly 
 on a logarithmic scale. Yet as we see the disagreement 
between theory and experiment is quite forbidding. With this 
remark we wish to demonstrate the need for 
comprehensive and precisely measured hadron abundance data sample,
including abundances of multi-strange antibaryons,
which were already 20 years ago identified as the 
hadronic signals of QGP phenomena \cite{Raf80}. The strange antibaryon
enhancement reported by the experiment WA97 fully confirms the
role played by these particles \cite{WA97}.

In  the approach  C, we introduce strangeness 
chemical non-equilibrium \cite{Raf91}, 
{\it i.e.}, we  also vary $\gamma_{s}$, 
keeping  $\gamma_q=1$.  A nearly valid 
experimental data description is now possible, 
and indeed, when the error bars were 
smaller  this was  a  satisfactory approach 
adapted in many studies \cite{Sol97}. 
However, only the  possibility of light quark 
non-equilibrium in fit D produces a statistically 
significant data description. 

It is  interesting to note that a significant 
degradation of  $\chi^2_{\rm T}$ occurs, 
especially in the  Pb--Pb data, when we require in column F 
that the particle ratios  comprising 
$\Omega$ and $\overline\Omega$-particles are 
also described. We thus conclude that a large fraction 
of these particles must be made in processes 
that are not considered in the present model. 

Another  notable  study case is shown in column D$_s$,
with strangeness conservation enforced. Remarkably, the
S--W data,  table~\ref{resultsw}, do not like this constraint. 
A possible explanation is that for S-induced reactions, 
the particle abundances are 
obtained at relatively high $p_\bot$\,. Thus only a small fraction of all
strange particles is actually observed, and therefore 
the overall strangeness is hard to balance.
Similar conclusion results also when radial 
flow is explicitly allowed for, with a significant 
unbalanced strangeness fractions remaining, 
as we shall discuss below. On the other hand,
this constraint is relatively easily satisfied 
for the Pb--Pb collision results,
table~\ref{resultpb}, where a much greater proportion 
of all strange particles is actually experimentally detected.

\begin{table}[t!]
\caption{\label{fitsw}
Statistical parameters which best describe the experimental  S--Au/W/Pb 
results  shown in table~\protect\ref{resultsw}. Asterisk ($^*$)  
means a fixed (input) value or result of a constraint.
In approaches B to D, particle abundance ratios comprising $\Omega$
are not considered. 
In case D$_{s}$  strangeness conservation in 
the particle  yields was enforced. 
In case F the three data-points with $\Omega$ are considered. 
Lower index $v$ implies 
that radial collective flow velocity has been considered. }
\vspace{-0.2cm}\begin{center}
\begin{tabular}{|l|cccccc|c|}
\hline\hline
{\small S--W}&$T_{\rm f}$ [MeV]& $\lambda_{q}$&$\lambda_{s}$&
$\gamma_{s}/\gamma_{q}$&$\gamma_{q}$&$v_c$& $\chi^2_{\rm T}$ \\
\hline
                    B
		     & 144 $\pm$ 2
                 & 1.53 $\pm$ 0.02
                 &   0.97 $\pm$ 0.02 
                 &   1$^*$
                 &   1$^*$
                 &   0$^*$
                 &   264  \\
                    C
		     &  147 $\pm$ 2 
                 & 1.49 $\pm$ 0.02
                 &   1.01 $\pm$ 0.02 
                 &   0.62 $\pm$ 0.02 
                 &   1$^*$
                 &   0$^*$
                 &   30  \\
                  D
		     & 143 $\pm$ 3 
                 & 1.50 $\pm$ 0.02
                 & 1.00 $\pm$ 0.02 
                 & 0.60 $\pm$ 0.02 
                 & 1.22 $\pm$ 0.06
                 &   0$^*$
                 & 6.5  \\
                    D$_s$
		     &  153 $\pm$ 3 
                 &   1.42 $\pm$ 0.02
                 &   1.10$^*\!\!$ $\pm$ 0.02 
                 &   0.56 $\pm$ 0.02 
                 &   1.26 $\pm$ 0.06
                 &   0$^*$
                 &   38  \\
                    F
		     &  144 $\pm$ 3 
                 &  1.49 $\pm$ 0.02
                 &  1.00 $\pm$ 0.02 
                 &  0.60 $\pm$ 0.02 
                 &  1.22 $\pm$ 0.06
                 &   0$^*$
                 &  12  \\
\hline
                 D$_v$
                 &  144 $\pm$ 2
                 & 1.51 $\pm$ 0.02
                 & 1.00 $\pm$ 0.02
                 & 0.69 $\pm$ 0.03
                 & 1.41 $\pm$ 0.08
                 & 0.49 $\pm$ 0.02
                 & 6.2\\
                 F$_v$
                 &  145$\pm$ 2
                 & 1.50 $\pm$ 0.02
                 & 0.99 $\pm$ 0.02
                 & 0.69 $\pm$ 0.03
                 & 1.43 $\pm$ 0.08
                 & 0.50 $\pm$ 0.02
                 & 11\\
\hline\hline
\end{tabular}
\end{center}
\end{table}
\begin{table}[t!]
\caption{\label{fitpb}
Statistical parameters which best describe the experimental
Pb--Pb results  shown in table~\protect\ref{resultpb}\,. Asterisk ($^*$)  
means a fixed (input) value, or result of a constraint.
In approaches B to D, particle abundance ratios comprising $\Omega$
are not considered. 
In case D$_{s}$  strangeness conservation in 
the particle  yields was enforced. 
In case F the four data-points with $\Omega$ are considered. Lower index $v$ implies 
that radial flow velocity has been considered.}
\vspace{-0.2cm}\begin{center}
\begin{tabular}{|l|cccccc|c|}
\hline\hline
{\small Pb--Pb}&$T_{\rm f} [MeV]$& $\lambda_{q}$&$\lambda_{s}$&
$\gamma_{s}/\gamma_{q}$&$\gamma_{q}$& $v_c$&$\chi^2_{\rm T}$\\
\hline
                    B
                 &  142 $\pm$ 3
                 &  1.70 $\pm$ 0.03
                 &  1.10 $\pm$ 0.02
                 &  1$^*$
                 &  1$^*$
                 &  0$^*$
                 &  88  \\

                    C
                 & 144 $\pm$ 4
                 & 1.62 $\pm$ 0.03
                 & 1.10 $\pm$ 0.02
                 & 0.63 $\pm$ 0.04
                 & 1$^*$
                 &  0$^*$
                 & 24  \\
                    D
                 &    134 $\pm$ 3
                 &   1.62 $\pm$ 0.03
                 &   1.10 $\pm$ 0.02
                 &   0.69 $\pm$ 0.08
                 &   1.84 $\pm$ 0.30
                 &   0$^*$
                 &    1.6       \\

                    D$_s$
                 &   133 $\pm$ 3
                 &   1.63 $\pm$ 0.03
                 &   1.09$^*\!\!$ $\pm$ 0.02
                 &   0.72 $\pm$ 0.12
                 &   2.75 $\pm$ 0.35
                 &  0$^*$
                 &   2.7       \\

                    F
                 &    334 $\pm$ 18
                 &   1.61 $\pm$ 0.03
                 &   1.12 $\pm$ 0.02
                 &   0.50 $\pm$ 0.01
                 &   0.18 $\pm$ 0.02
                 &  0$^*$
                 &   19       \\
\hline
                 D$_v$
                 &  144 $\pm$ 2
                 & 1.60 $\pm$ 0.02
                 & 1.10 $\pm$ 0.02
                 & 0.86 $\pm$ 0.03
                 & 1.72 $\pm$ 0.08
                 & 0.58 $\pm$ 0.02
                 & 1.5\\
                 F$_v$
                 &  328 $\pm$ 17
                 & 1.59 $\pm$ 0.03
                 & 1.13 $\pm$ 0.02
                 & 0.51 $\pm$ 0.03
                 & 0.34 $\pm$ 0.13
                 & 0.38 $\pm$ 0.31
                 & 18\\
\hline\hline
\end{tabular}
\end{center}
\end{table}

The statistical parameters associated with the 
particle abundances described above 
are shown in the table~\ref{fitsw} for S-W/Pb 
and in table~\ref{fitpb} for  Pb--Pb 
reactions. The errors of the statistical parameters 
shown are those provided  by the program MINUIT96.03 from CERN
program library. When the theory describes 
the data well, this is a one standard 
deviation error in theoretical parameters arising 
from the experimental measurement 
error. In tables~\ref{fitsw}~and~\ref{fitpb} in cases in which $\gamma_q\ne 1$ 
we present the ratio $\gamma_s/\gamma_q$, which corresponds approximately 
to the parameter $\gamma_s$ in the data
studies in which $\gamma_q=1$ has been  assumed. It is notable that whenever 
we allow  phase space occupancy to vary 
from the equilibrium, a significant deviation is found.
In the S--W case, table~\ref{fitsw}, there 
is a 25\% excess in the light quark occupancy, 
while strange quarks are 25\% below equilibrium.
In Pb--Pb case, table~\ref{fitpb}, the 
ratio of the nonequilibrium parameters
$\gamma_s/\gamma_q\simeq 0.7$ also varies little 
(excluding the failing cases F with
$\Omega,\overline\Omega$ data), though 
the individual values $\gamma_s, \gamma_q$
can change significantly, even  between the 
high confidence cases. 

We note that in the Pb--Pb reaction, table~\ref{fitpb}, $\gamma_s>1$.
This is an important finding, since an explanation of this effect 
involves formation prior to freeze-out in the matter at high density of near 
chemical equilibrium, $\gamma_s(t<t_f)\simeq 1$. The ongoing  rapid expansion 
(note that the collective velocity at freeze-out is found to  be $1/\sqrt{3}$)
preserves this high strangeness yield, and thus we find the result 
$\gamma_s>1$. In other words the strangeness production 
freeze-out temperature $T_s>T_f$. Thus the strangeness equilibration time
is proven implicitly to be of magnitude expected in earlier studies of the 
QGP processes \cite{acta96}. It is hard, if not really impossible, to 
arrive at this result  without the QGP hypothesis. Moreover, inspecting
figure 38 in \cite{acta96} we see that the yield of strangeness we expect
from the kinetic theory in QGP is at the level of 0.75 per baryon, the level
we indeed will determine below.

Another notable results is 
$\tilde\lambda_s\simeq\lambda_{s}\simeq 1.0$ 
in the S--Au/W/Pb case, see table \ref{fitsw}, 
and $\lambda_{s}\simeq 1.1$ in the Pb--Pb case, see table \ref{fitpb}, 
implying here  $\tilde\lambda_{s}=1$, see section \ref{Coulombsec}. 
We see clearly  for both S- and Pb-induced reactions a value of $\lambda_s$,
characteristic for a source of freely movable 
strange quarks with balancing strangeness.

\section{Physical properties of the fireball at  chemical freeze-out} 
Given the precise statistical information about the properties of the 
hadron phase space, we can determine the physical properties
of the hadronic particles, see tables~\ref{tqsw}~and~\ref{tqpb}.
We show for the same study cases B--F, along with their temperature, the
specific energy and entropy content per baryon, 
and specific anti-strangeness content,
along with specific strangeness asymmetry, and finally pressure at 
freeze-out. We note that it is improper in general to refer to these properties 
as those of a `hadronic gas' formed in nuclear collisions, as the particles 
considered may be emitted in sequence from a deconfined source, and thus there 
may never be a evolution  stage corresponding to a 
hadron gas phase. However, the properties presented
are those carried away by the emitted particles, 
and thus characterize the properties of their source.

\begin{table}[tb]
\caption{\label{tqsw}
$T_{\rm f}$ and physical properties 
(specific energy, entropy, anti-strangeness, net strangeness,
pressure and volume) of  the full 
hadron phase space characterized by the statistical parameters  
given in table~\protect\ref{fitsw} for the reactions S--Au/W/Pb. 
Asterisk~$^*$ means fixed input.} 
\vspace{-0.2cm}\begin{center}
\begin{tabular}{|l|cccccc|} 
\hline\hline
{\small S--W}&$T_{\rm f}$ [MeV]& $E_{\rm f}/B$  & $S_{\rm f}/B$ & ${\bar s}_{\rm f}/B$ 
& $({\bar s}_{\rm f}-s_{\rm f})/B$ &$P_{\rm f}$ [GeV/fm$^3$] \\
\hline
                    B
		     & 144 $\pm$ 2
                 & 8.9 $\pm$ 0.5
                 &  50$\pm$ 3
                 &  1.66 $\pm$ 0.06
                 &  0.44 $\pm$ 0.02  
                 &  0.056 $\pm$ 0.005 \\
                    C
		     &  147 $\pm$ 2
                 & 9.3 $\pm$ 0.5
                 & 49$\pm$ 3
                 & 1.05 $\pm$ 0.05
                 & 0.23 $\pm$ 0.02  
                 & 0.059 $\pm$ 0.005   \\
                    D
		     &  143 $\pm$ 3
                 & 9.1 $\pm$ 0.5
                 & 48$\pm$ 3
                 & 0.91 $\pm$ 0.04
                 & 0.20 $\pm$ 0.02  
                 & 0.082 $\pm$ 0.006  \\ 
                    D$_s$
		     &   153 $\pm$ 3
                 &  8.9 $\pm$ 0.5
                 &  45 $\pm$ 3
                 &  0.76 $\pm$ 0.04
                 &  0$^*$  
                 &  0.133 $\pm$ 0.008 \\ 
                    F
		     &  144 $\pm$ 2
                 & 9.1 $\pm$ 0.5
                 & 48 $\pm$ 3
                 & 0.91 $\pm$ 0.05
                 & 0.20 $\pm$ 0.02  
                 & 0.082 $\pm$ 0.006   \\
\hline
                    D$_v$
		     &   144 $\pm$ 2
                 &  8.2 $\pm$ 0.5
                 &  44 $\pm$ 3
                 &  0.72 $\pm$ 0.04
                 &  0.18 $\pm$ 0.02  
                 &  0.124 $\pm$ 0.007 \\ 
                    F$_v$
		     &  145 $\pm$ 2
                 & 8.2 $\pm$ 0.5
                 & 44 $\pm$ 3
                 & 0.73 $\pm$ 0.05
                 & 0.17 $\pm$ 0.02  
                 & 0.123 $\pm$ 0.007   \\
\hline\hline
\end{tabular}
\end{center}
\end{table}
\begin{table}[tb]
\caption{\label{tqpb}
$T_{\rm f}$ and physical properties 
(specific energy, entropy, anti-strangeness, net strangeness,
pressure and volume) of  the full 
hadron phase space characterized by the statistical parameters  
given in table~\protect\ref{fitpb} for the reactions Pb--Pb. 
Asterisk~$^*$ means fixed input.}
\vspace{-0.2cm}\begin{center}
\begin{tabular}{|l|lccccc|}
\hline\hline
{\small Pb--Pb}&$T_{\rm f}$ [MeV]& $E_{\rm f}/B$  & $S_{\rm f}/B$ & ${\bar s}_{\rm f}/B$ & 
$({\bar s}_{\rm f}-s_{\rm f})/B$ &$P_{\rm f}$ [GeV/fm$^3$] \\
\hline
                    B
                 & 142 $\pm$ 3
                 & 7.1 $\pm$ 0.5
                 &  41 $\pm$ 3
                 &  1.02 $\pm$ 0.05
                 &  0.21 $\pm$ 0.02
                 &  0.053 $\pm$ 0.005 \\
                    C
                 &  144 $\pm$ 4
                 & 7.7 $\pm$ 0.5
                 & 42 $\pm$ 3
                 & 0.70 $\pm$ 0.05
                 & 0.14 $\pm$ 0.02
                 & 0.053 $\pm$ 0.005 \\
                    D
                 &  134 $\pm$ 3
                 &  8.3 $\pm$ 0.5
                 &  47 $\pm$ 3
                 &  0.61 $\pm$ 0.04
                 &  0.08 $\pm$ 0.01
                 &  0.185 $\pm$ 0.012 \\
                    D$_s$
                 &   133 $\pm$ 3
                 &  8.7 $\pm$ 0.5
                 &  48 $\pm$ 3
                 &  0.51 $\pm$ 0.04
                 &  0$^*$
                 &  0.687 $\pm$ 0.030 \\
                    F
                     &  334 $\pm$ 18
                 & 9.8 $\pm$ 0.5
                 & 24 $\pm$ 2
                 & 0.78 $\pm$ 0.05
                 & 0.06 $\pm$ 0.01
                 & 1.64 $\pm$ 0.06 \\
\hline
                    D$_v$
		     &   144 $\pm$ 2
                 &  7.0 $\pm$ 0.5
                 &  38 $\pm$ 3
                 &  0.78 $\pm$ 0.04
                 &  0.01 $\pm$ 0.01  
                 &  0.247 $\pm$ 0.007 \\ 
                    F$_v$
		     &  328 $\pm$ 17
                 & 11.2 $\pm$ 1.5
                 & 28 $\pm$ 3
                 & 0.90 $\pm$ 0.05
                 & 0.09 $\pm$ 0.02  
                 & 1.40 $\pm$ 0.06   \\
\hline\hline
\end{tabular}
\end{center}
\end{table}
The energy  per baryon seen in the emitted hadrons is, within error, equal
to the available specific energy  of the collision  (8.6 GeV for Pb--Pb and about 8.8--8.9 GeV
for S--Au/W/Pb). This implies that the fraction of energy left in the central 
fireball  must be the same as the fraction of baryon number.
We further note that hadrons emitted at freeze-out carry away a 
specific $\bar s$ content which is determined to be $0.72\pm0.04$
in S--Au/W/Pb case and $0.78\pm0.04$ for the
Pb--Pb collisions (cases D$_v$). Here we see the most significant 
impact of flow, as without it the specific strangeness content 
seemed to diminish as we moved to the larger collision system. 
We have already alluded repeatedly to the fact that for S--Au/W/Pb 
case the balance of strangeness is not seen in the particles observed 
experimentally. The asymmetry is 18\%, with the 
excess  emission of $\bar s$ containing hadrons at high $p_\bot>1$ GeV. 
In the Pb--Pb data this effect disappears, perhaps since the $p_\bot$
lower cut-off is smaller. One could also imagine that
longitudinal flow which is stronger in S--Au/W/Pb is responsible 
for this effect. 

The small reduction of the specific entropy in Pb--Pb compared to 
the lighter S--Au/W/Pb is driven by the greater baryon 
stopping in the larger system, also seen in the smaller energy per
baryon content. Both systems freeze out at $E/S=0.185$ GeV (energy per unit
of entropy). Aside of $T_f$, this  is a second universality feature at
hadronization of both systems. 
The overall high specific entropy content agrees well with the entropy content 
evaluation made  earlier \cite{Let93} for the S--W case.
 This is so because the strange particle data are indeed 
fully consistent within the chemical-nonequilibrium description  with the $4\pi$
total particle multiplicity results. 

\section{Current status and conclusions}\label{secsum}

We have presented detailed analysis of hadron abundances observed
in central S--Au/W/Pb 200 A GeV and Pb--Pb 158 A GeV interactions
within thermal equilibrium and chemical non-equilibrium  phase 
space model of strange and non-strange hadronic particles.
In the analysis of the freeze-out the structure of  the 
particle source was irrelevant. However, the results 
that we found for the statistical parameters point rather clearly towards 
a deconfined QGP source, with quark abundances near but not at
chemical equilibrium: statistical parameters
obtained characterize a strange particle source which, both for 
S--Au/W/Pb and for  Pb--Pb case, when allowing for Coulomb deformation 
of the strange  and anti-strange quarks,
is exactly symmetric between $s$ and $\bar s$ quark carriers, 
as is natural for a deconfined state.

There are similarities in freeze-out properties seen comparing 
results presented in tables~\ref{tqsw}~and~\ref{tqpb}
which suggest universality in
the extensive physical properties of the two freeze-out 
systems \cite{LRPb98}. Despite considerably varying statistical parameters
we obtain similar physical conditions such as $T_f$ and $E/S$
 corresponding possibly to the common
physical properties of QGP at its breakup into hadrons.
The precision of the data description we have reached, see 
tables~\ref{resultsw}--\ref{Tetrange} strongly suggests that 
despite its simplicity the model we developed to analyze experimental
data provides a reliable image of the hadron production processes. 
When flow is  allowed for
the freeze-out temperature is identical in both physical systems considered, 
even though, {\it e.g.,} the baryochemical potential
 $\mu_B=3T\ln\lambda_q$  and other physical parameters found are 
slightly different. It is worth to restate the values we obtained, case D$_v$ 
(6 parameter description, no $\Omega,\,\overline\Omega$):
\[T_f=144\pm2\,\mbox{MeV},\quad \mu_B=178\pm5\mbox{MeV\,,\quad for\ S--Pb/W}\,,\]
\[T_f=144\pm2\,\mbox{MeV},\quad \mu_B=203\pm5\mbox{MeV\,,\quad for\ Pb--Pb}\,,\ \ \ \]

 Even though
there is still considerable uncertainty about other freeze-out 
flow effects, such as longitudinal
flow (memory of the collision axis), the level of consistency
and quality of agreement between a wide range of experimental data
and our chemical non-equilibrium, thermal equilibrium statistical 
model suggests that, for the observables considered, 
these effects do not matter.

The key results we found analyzing experimental data  are:\begin{enumerate}
\item $\tilde \lambda_s=1$ for
S and Pb collisions\,;
\item $\gamma_s^{\rm Pb}>1, \ \gamma_q>1$\,;
\item $v_c^{\rm Pb}=1/\sqrt{3}$
\item $S/B\simeq 40$\,;
\item $\bar s/B\simeq 0.75$\,;
\end{enumerate}
are in remarkable agreement with the properties of a deconfined 
QGP source hadronizing  without chemical reequilibration. 
The only natural interpretation of our findings is thus that hadronic
particles seen at 158--200 GeV A nuclear collisions at CERN-SPS
are  emerging directly from a deconfined QGP state
and do not undergo a chemical re-equilibration after they have been 
produced.

{\vspace{0.5cm}\noindent\it Acknowledgments:\\}
This work was supported in part
by a grant from the U.S. Department of
Energy,  DE-FG03-95ER40937\,. LPTHE, Univ.\,Paris 6 et 7 is:
Unit\'e mixte de Recherche du CNRS, UMR7589.


\end{document}